\documentclass[12pt,aps]{revtex4}
\usepackage{epsfig}

\def\bea{\begin{eqnarray}}
\def\eea{\end{eqnarray}}
\def\nn{\nonumber}

\def\O{{\cal O}}
\begin{document}

\title{Distinguishing among Scalar Field Models of Dark Energy }

\author{Irit Maor, Ram Brustein}

\address{ Department of Physics, Ben-Gurion University,
Beer-Sheva 84105, Israel\\
\texttt{e-mail:  irrit,ramyb@bgumail.bgu.ac.il }}

\begin{abstract}

We show that various scalar field models of dark energy predict
degenerate luminosity distance history of the Universe and thus
cannot be distinguished by supernovae measurements alone. In
particular, models with a vanishing cosmological constant (the value
of the potential at its minimum) are degenerate with models with a
positive or negative cosmological constant whose magnitude can be as
large as the critical density. Adding information from CMB anisotropy
measurements does reduce the degeneracy somewhat but not
significantly. Our results indicate that a theoretical prior on the
preferred form of the potential and the field's initial conditions
may allow to quantitatively estimate model parameters from data.
Without such a theoretical prior only limited qualitative information
on the form and parameters of the potential can be extracted even
from very accurate data.
 \end{abstract}
\maketitle

\section{Introduction}

One of the standard methods of interpreting the growing body of
evidence  from supernovae \cite{data} and other measurements  that
the expansion of the Universe is accelerating, is to assume the
existence of a dark energy component and to model it using scalar
fields (for a recent review see \cite{fund}). This links the
expansion history of the Universe to theories of fundamental
physics. For example, from this perspective the value of the
potential at its minimum is the cosmological constant (CC). Since
at this point there are many theoretical ideas about the form of
the potential but none that particularly stands out, it would
have been helpful if the data from cosmological measurements,
such as supernovae Ia, CMB  and various others, could provide
hints about some generic features of the potential.

Viable scalar field models of dark energy need to have potentials
whose energy scale is about the critical density $\sim
10^{-12}{\rm eV}^4$, and typical scalar field masses about the
Hubble mass $m\sim 10^{-33}$ eV. In such models typical scalar
field variations are about Planck scale $m_p\sim 10^{19}$ GeV,
and typical time scales for such variations are about the Hubble
time $1/H_0\sim 10^{18}$ sec. Whether, and how well, it is
possible to determine the parameters and form of potentials and
the field's dynamical history and future from data beyond such
qualitative estimates has been addressed previously
\cite{Chiba,Barger,weller,Ng,Bludman,Eriksson,hut,star1,efst}.
Weller and Albrecht \cite{weller} concluded that some potentials
could be differentiated using SNAP-like data \cite{snap}. They
approximated the equation of state (EOS) of each of the models,
and showed by likelihood analysis that some of the models are
distinguishable. Another approach is ``reconstruction"
\cite{hut,star1,efst}. Here one rewrites the potential as a
function of the luminosity-distance ($d_L$) and its derivatives,
which are in turn functions of the redshift. The potential (and
not the EOS) is approximated by a fitting function, and
statistically tested against a set of accurate $d_L$
measurements. The efficiency of reconstruction depends on the
accuracy of knowing the values of $\Omega_m$ (this is needed also
when one fits the EOS), and $H_0$ (this is needed only for
reconstruction).

On general grounds we expect that scalar field potentials are
less distinguishable than their corresponding EOS, because
different potentials, with properly adjusted initial conditions
for the field can produce very similar EOS.

In previous papers \cite{m1,m2} we have found that supernovae (SN) 
measurements are limited as a probe of the dark energy EOS $w_Q$, due
to degeneracies. Specifically, it was shown that $d_L$'s
corresponding to two different $w_Q$'s are degenerate if both EOS
coincide at some point at a relatively low red shift, $z^*$ (see also
\cite{sweetspot} and \cite{astier}). The purpose of the present
analysis is to explore the implications of this degeneracy on the
possibility to determine the scalar field potential. For a given
functional form of potential, we would like to quantify the amount by
which the parameters of the potential can be varied, and still be
indistinguishable by accurate SN measurements. Our criteria for
indistinguishability between two models is that their resulting
$d_L$'s differ at most by 1\% up to redshift $z=2$, in accordance
with the anticipated accuracy of future SN measurements.  In
addition, we would like to determine whether the functional form of
the potential can be distinguished or constrained by data.

We look for degeneracies among potentials using the following
procedure: For a given class of potentials, we change the
parameters as well as the initial values of the field, with the
constraint that $w_Q$ at $z^*$ remains unchanged. This results in
models whose $w_Q$ cross at $z^*$. We know from our previous
analysis that in this case the models tend to be degenerate. Then
we evaluate numerically the differences in the $d_L$'s of the models
to verify this.

There are additional sources of degeneracy that we do not consider
here. In our procedure  the value of the potential energy and the
value of the kinetic energy remain unchanged. Allowing changes in the
potential that are compensated by changes in the initial
conditions for the kinetic energy will give another dimension of
degeneracy. Variation in the value of $\Omega_m$ is yet another
degree of freedom, as is relaxing the assumption of a flat
Universe and considering the effects of a clumpy
Universe \cite{sps}. Additionally, the value of $z^*$ in different
models can be shifted. We have found that $z^*$ varies slowly
with the red shift depth of the data set, $z_{max}$. The
dependence is approximately linear $z^*=\alpha z_{max}+\beta$
(see also \cite{linder}), $\alpha$ and $\beta$ are model
dependent but $\alpha$ is typically small, about 0.2. And,
finally, we have looked only at a class of simple potentials that
have two independent parameters. Additional parameters in the
potential yet again open up new degrees of freedom, each of which
increases the degeneracy of each of the parameters. Since we have
found that this class of simple potentials suffers from large
degeneracies, we see no phenomenological justification for using
more complex potentials. If a theoretical prior about the form of
the potential and initial conditions can be motivated then some
of this degeneracy can be removed.

\section{Degeneracies of scalar field potentials} \label{deg}

We consider a flat Universe with non-relativistic matter (dark
matter included) whose EOS is $w_m=0$, and a dark energy
component which we model by a canonically normalized and
minimally coupled scalar field. Einstein's equations for such a
Universe are the following,
  \bea
 & H^2=\kappa^2\left(\rho_m x^3+\frac{1}{2}x^2H^2\phi'^2
   +V(\phi)\right ) & \label{cons} \\
 & xHH'=\frac{3\kappa^2}{2}\left(\rho_m x^3+x^2H^2\phi'^2 \right) &  \\
 & x^ 2H^2\phi''+(x^2HH'-2xH^2)\phi'+\frac{dV}{d\phi}=0, &
 \eea
where $x=1+z$, $\kappa^2=\frac{8\pi}{3 m_p^2}$, $\rho_m$ is the
matter energy density today, primes denote derivatives with
respect to $x$, and $V$ is the potential of the scalar field. The
scalar field's EOS ($w_Q$) is given by,
  \bea
   w_Q&=&\frac{p_Q}{\rho_Q}\nonumber \\
  &=&\frac{x^2H^2\phi'^2-2V}{x^2H^2\phi'^2+2V}.   \label{eos}
\eea

We choose a model, and vary the parameters of the potential
$P_i$, as well as the values of the scalar field at $z^*$ keeping
its derivative constant,
  \bea
     \phi(z^*)&\rightarrow& \phi(z^*)+\delta\phi  \\
     V &\rightarrow& V+\delta V(P_i,\delta P_i,~\delta \phi)  \\
     \delta\phi'(z^*)&=&\delta H(z^*)=\delta \rho_m=0.
 \eea
These variations result in variations in eqs.(\ref{eos}) and
(\ref{cons}):
  \bea
    w_Q+\delta w_Q&=&\frac{x^2H^2\phi'^2-2(V+\delta V)}
                 {x^2H^2\phi'^2+2(V+\delta V)}\nonumber \\
         &\approx&
                 w_Q+\left[\frac{2(1+w_Q)}{x^2H^2\phi'^2+2V}
                 \right]\delta V+\O(\delta V^2)  \label{vw} \\
     H^2&=&\kappa^2\left(\rho_m x^3+\frac{1}{2}x^2H^2\phi'^2
                 +(V(\phi)+\delta V)\right ).  \label{vc}
 \eea
In eq.(\ref{vw}), we have  kept terms up to first order in $\delta
V$, assuming it is small enough.

Since we know that if the values of $w_Q(z^*)$ are equal for two
models then their luminosity-distance history is approximately
degenerate, we explore part of the degeneracy in parameter space
by requiring that  $\delta w_Q(z^*)$ vanishes. In eq.~(\ref{vw})
we ignore higher orders in $\delta V$, so we simply require that
$\delta V$ vanishes to first order. On the other hand, even a
small deviation from a spatially flat cosmology will be amplified
by the evolution of the solution. We therefore need to require
that eq.~(\ref{vc}) holds exactly.

The two constraints we are imposing are
then
  \bea
\label{algebra1}
  & \delta V(P_i,\delta P_i,~\delta \phi)^{(1)}=0 & \\
  & \delta V(P_i,\delta P_i,~\delta \phi)^{(exact)}=0 .&
\label{algebra2}
 \eea
This set of variations and constraints is algebraic and can often be
solved analytically. The analytical solution connects any given
model to a family of associated models, all of which have the
same $w_Q$ at $z^*$. The next step is to check numerically how
large are the allowed parameter variations such that different
models in the family are indistinguishable. \footnote{Recall that
our criterion for indistinguishable models is that their $d_L$'s
do not differ by more than 1\% in the range $0\le z \le 2$.}

Initial conditions  are given at  $z^*$, but the evolution of
each of the models towards $z=0$ is different, and they end up
with a different value of $H_0$ (Hubble parameter at $x=1$,
$z=0$). To ensure that all models have the same value of $H_0$ we
rescale them,
  \bea
& \widetilde{H}=\frac{H}{H_0},~~\Omega_m=\frac{\rho}{H_0^2m_p^2},
               ~~\widetilde{\phi}=\frac{\phi}{m_p},
               ~~\widetilde{V}=\frac{V}{H_0^2m_p^2}.
 \eea
Equation~(\ref{cons}) can be reexpressed in terms of the rescaled
variables
 \bea
 & \widetilde{H}^2=\Omega_m x^3+\frac{1}{2}x^2\widetilde{H}^2
               \widetilde{\phi}'^2+\widetilde{V}.  &
 \eea
This means that although initially only the scalar field potential
energy is varied, eventually all quantities (except $H_0$) may
vary between models. As we shall show later (section \ref{r}), it
turns out that differences in $\Omega_m$ are negligible. This
result is not surprising: as was shown in \cite{m2} (see in
particular Fig.~2), this type of degeneracy characterizes models
with fixed values of $\Omega_m$. Families of models which are
degenerate with respect to SN measurements but have different
values of $\Omega_m$ do not exhibit the enhanced resolution of
$w_Q(z^*)$. This means that on top of the degeneracy that we will
be exploring here, another dimension of degeneracy will open up
once the uncertainties in $\Omega_m$ are taken into account. The
method described here  therefore exposes only part of the
degeneracy among potentials.

\section{Degeneracies among simple potentials} \label{r}

We have analyzed various forms of potentials with two parameters.
For all such potentials there are three independent variations, two
parameters and the initial condition for $\phi$. Enforcing the two
constraints eqs.~(\ref{algebra1}),(\ref{algebra2}) results in a
solution which is a curve in a three-dimensional space. Obviously the
degeneracy is larger when more parameters are allowed.

One  class of potentials we have looked at is
$V(\phi)=\frac{1}{2}m^2\phi^2+v_0$. This is one of the standard
simple forms of potentials, $m$ is the scalar field mass, $v_0$ is
the value of the potential at the minimum (the CC). As explained
previously  we already have an order of magnitude estimates for
$v_0$ and $m$, but we would  like to know whether they can be
determined in a more quantitative and conclusive way by the data.
In particular we would like to know whether it is possible to
distinguish models with a vanishing CC ($v_0=0$) from models with
a non-vanishing CC ($v_0\ne 0$). \footnote{It is worth noting here
that although $v_0\neq 0$ is interpreted as a contribution to the
vacuum energy which has $w_{\Lambda}=-1$, offsetting $v_0$ from
zero and allowing $V$ to have negative regions can yield
$|w_Q|>1$ while the field is still dynamical. For example, if
$V=-KE/3$ ($KE$ is the kinetic energy of the field), then
$w_Q=(KE-V)/(KE+V)=2$.} For this class of quadratic potentials
the variation is given by: \bea
     V+\delta V&=&\frac{1}{2}(m+\delta m)^2(\phi+\delta \phi)^2+
                 (v_0+\delta v_0) \nn \\
    &=&\left[\frac{1}{2}m^2\phi^2+v_0 \right]+
                 \left[m^2\phi\delta\phi+
                 m\phi^2\delta m+\delta v_0 \right]   \nn \\
    &+&\left[\frac{1}{2}(m\delta\phi+\phi\delta m)^2
                 +\delta\phi \delta m(m+\delta m)(\phi+\delta\phi)
                 -\frac{1}{2}\delta m^2 \delta \phi^2\right].
 \eea
The first bracket is $V$ itself. The second bracket is $\delta
V^{(1)}$ which should vanish, and the last bracket is $\delta
V^{(exact)}-\delta V^{(1)}$ which should vanish as well. The
solution is a curve in the $(\delta v_0, ~\delta m, ~\delta\phi)$
space. On this curve changes in the curvature and the height of
the minimum of the potential are compensated by a change in the
value of $\phi$ such  that the potential energy of the field is
locally unchanged.

\begin{figure}
  \begin{center}
    \epsfig{file=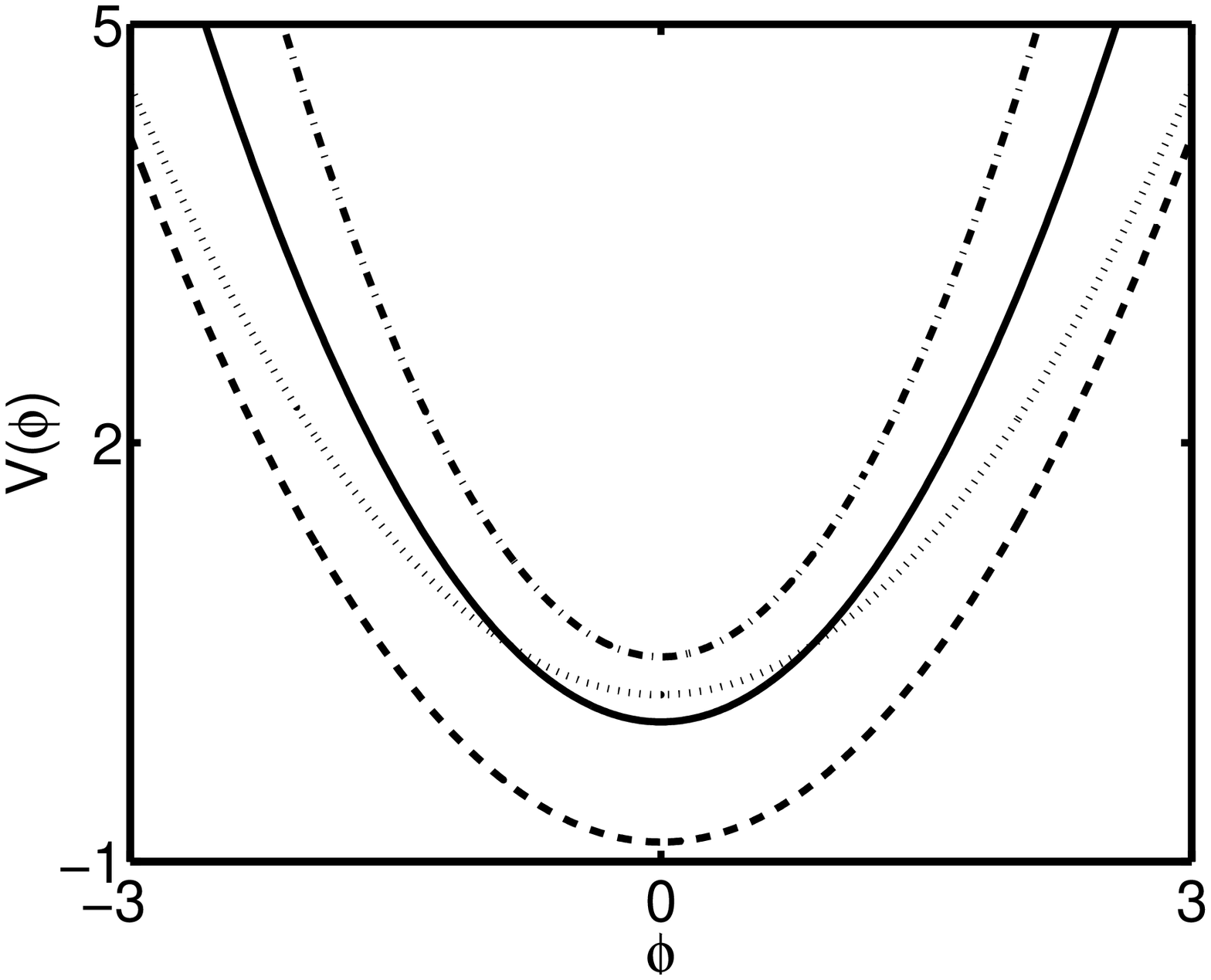,height=35mm}
    \epsfig{file=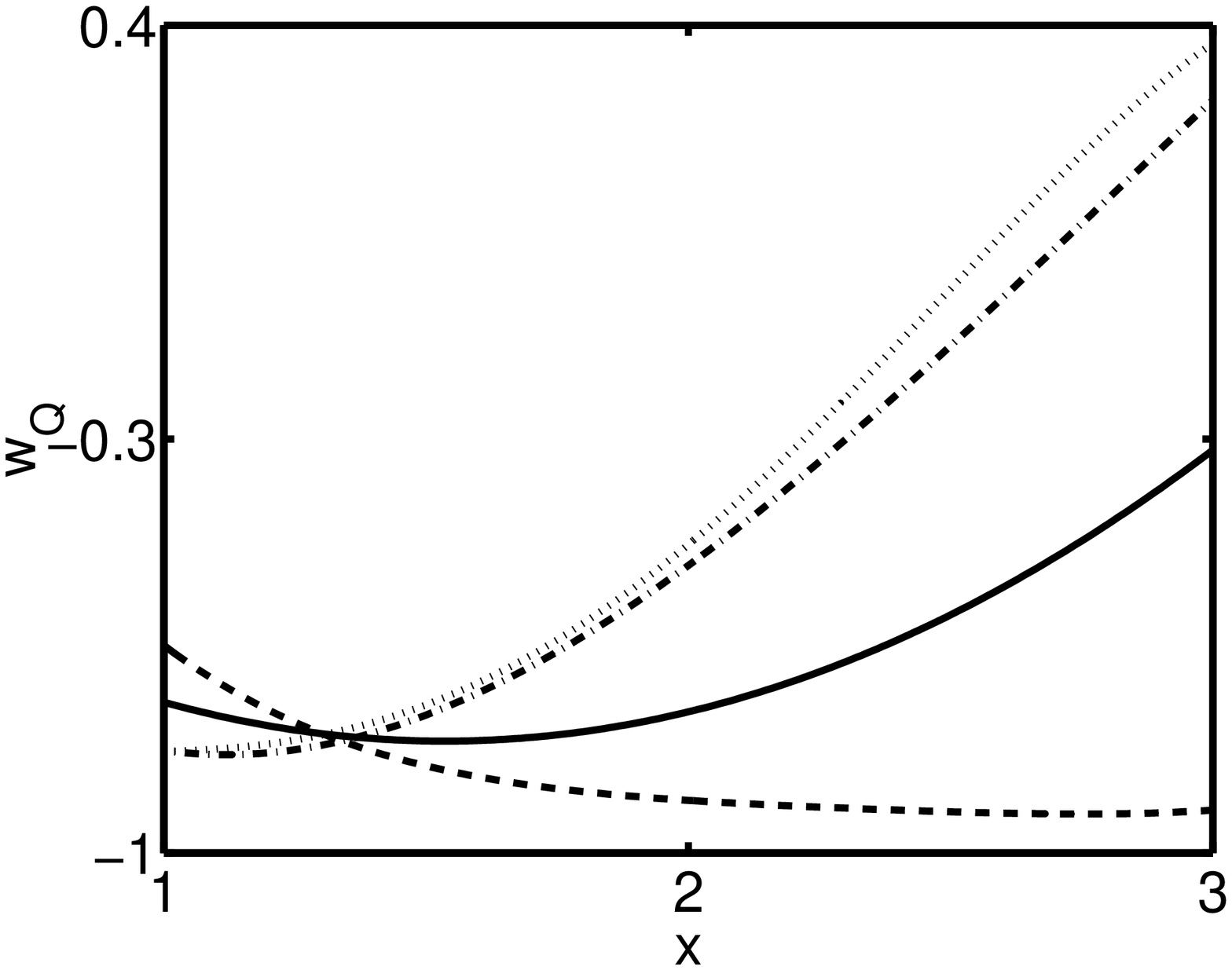,height=35mm}
    \epsfig{file=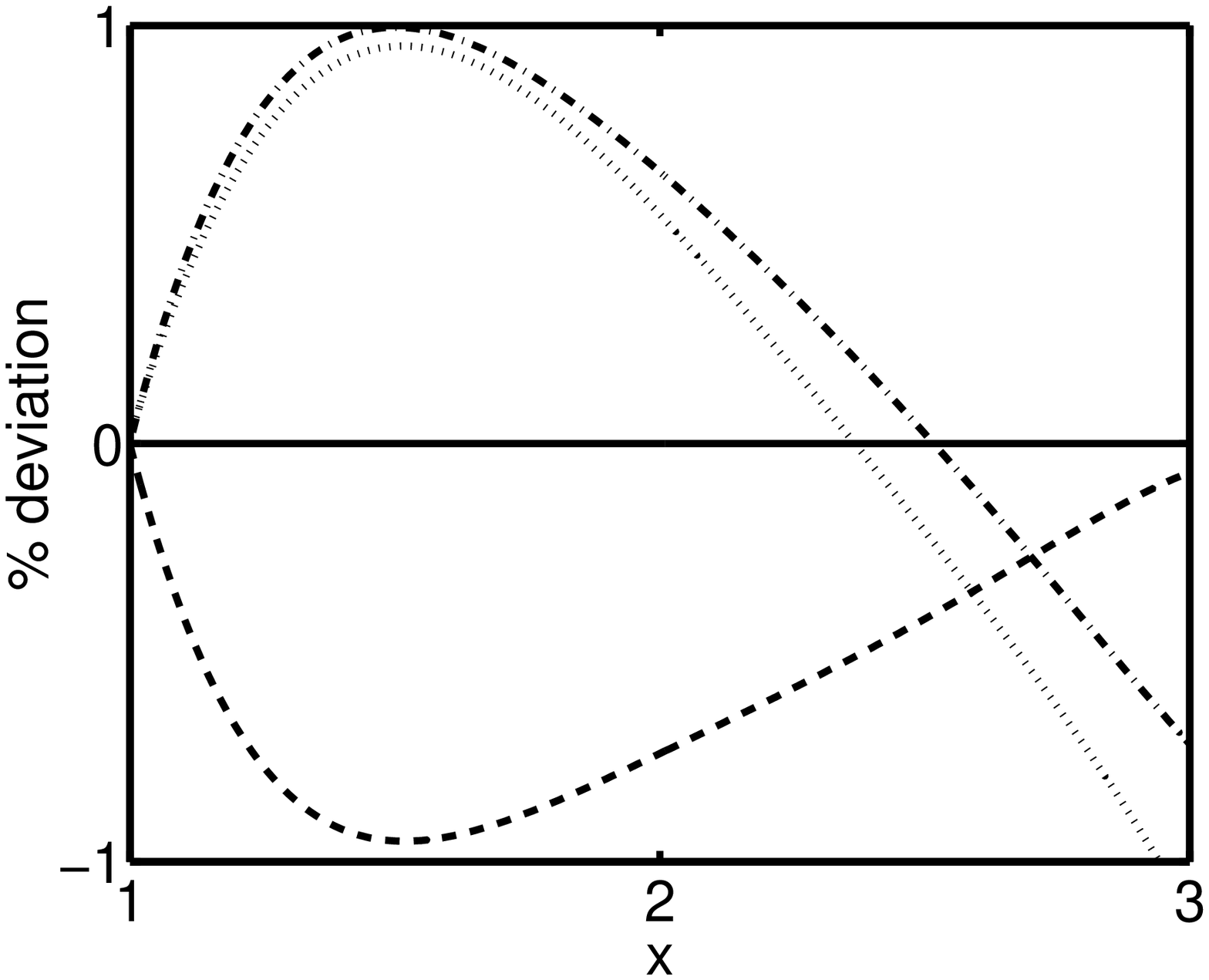,height=35mm}
  \end{center}
 \caption{Various quadratic potentials (shown on the left), their
           resulting $w_Q$ (middle), and the relative difference in
           $d_L$ (right). The parameters of the
           potentials are listed in the table I.}
  \label{m+}
 \end{figure}

Figure~\ref{m+} shows a variety of quadratic potentials on the left,
their resulting $w_Q$ in the middle, and the relative difference
in $d_L$ on the right. The parameters of the
potentials are listed in table I. Notice that although
the fiducial model has a vanishing CC,
$v_0=0$, it is degenerate with models that have $v_0$ of order
unity in units of the critical density ($m_p^2H_0^2$). The
uncertainty in $m$ is of order unity  in units of the
present value of the Hubble parameter  $H_0$. \\
All the models have values of $\Omega_m$ which are within $\pm
0.01$ from the value of the fiducial. This means that we are
indeed exploring here the degeneracy due to the integral relation
between $d_L$ and $w_Q$, and not the degeneracy due to the
uncertainty in $\Omega_m$. As explained previously, this
is a direct consequence of our method.

\begin{figure}
  \begin{center}
    \epsfig{file=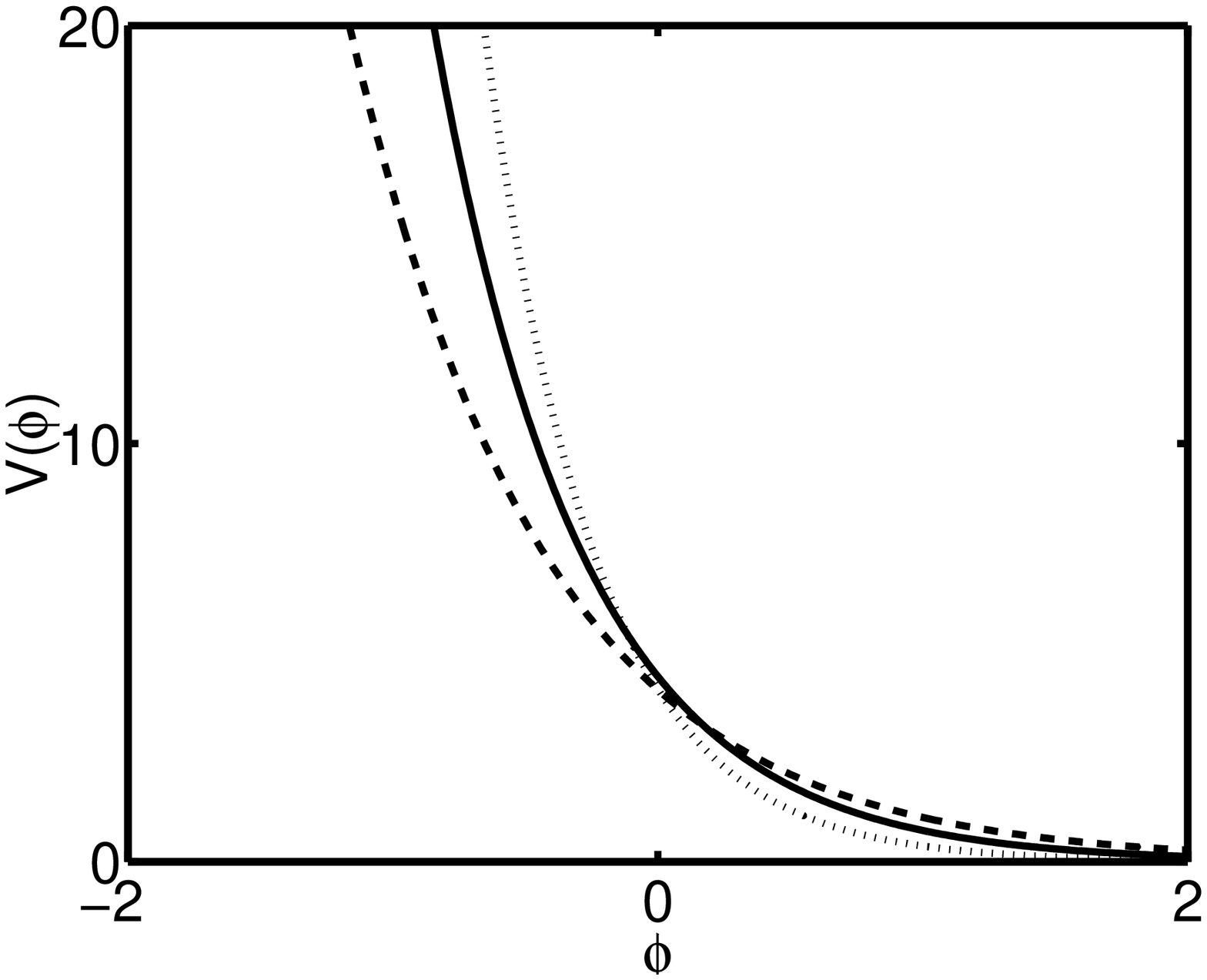,height=35mm}
    \epsfig{file=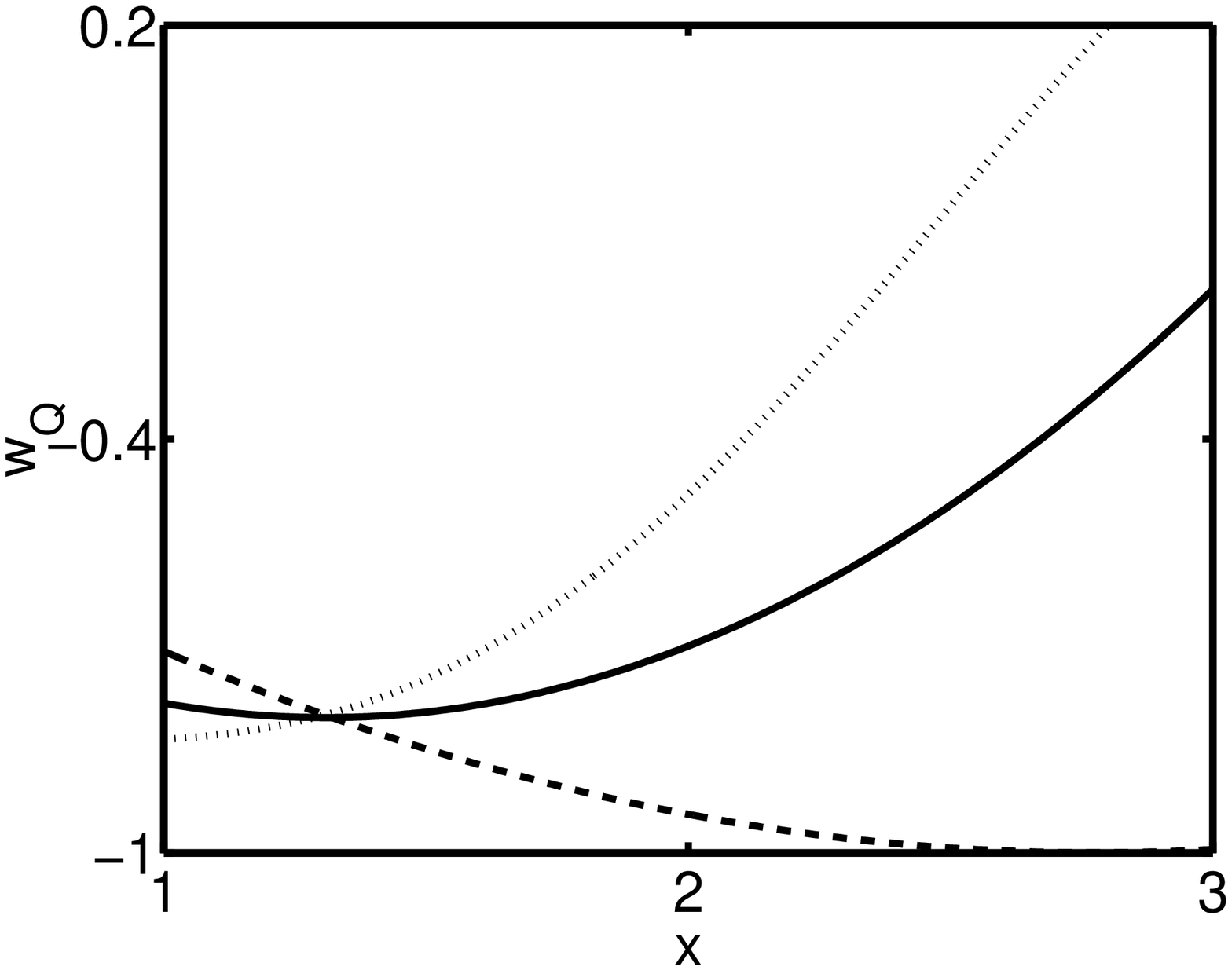,height=35mm}
    \epsfig{file=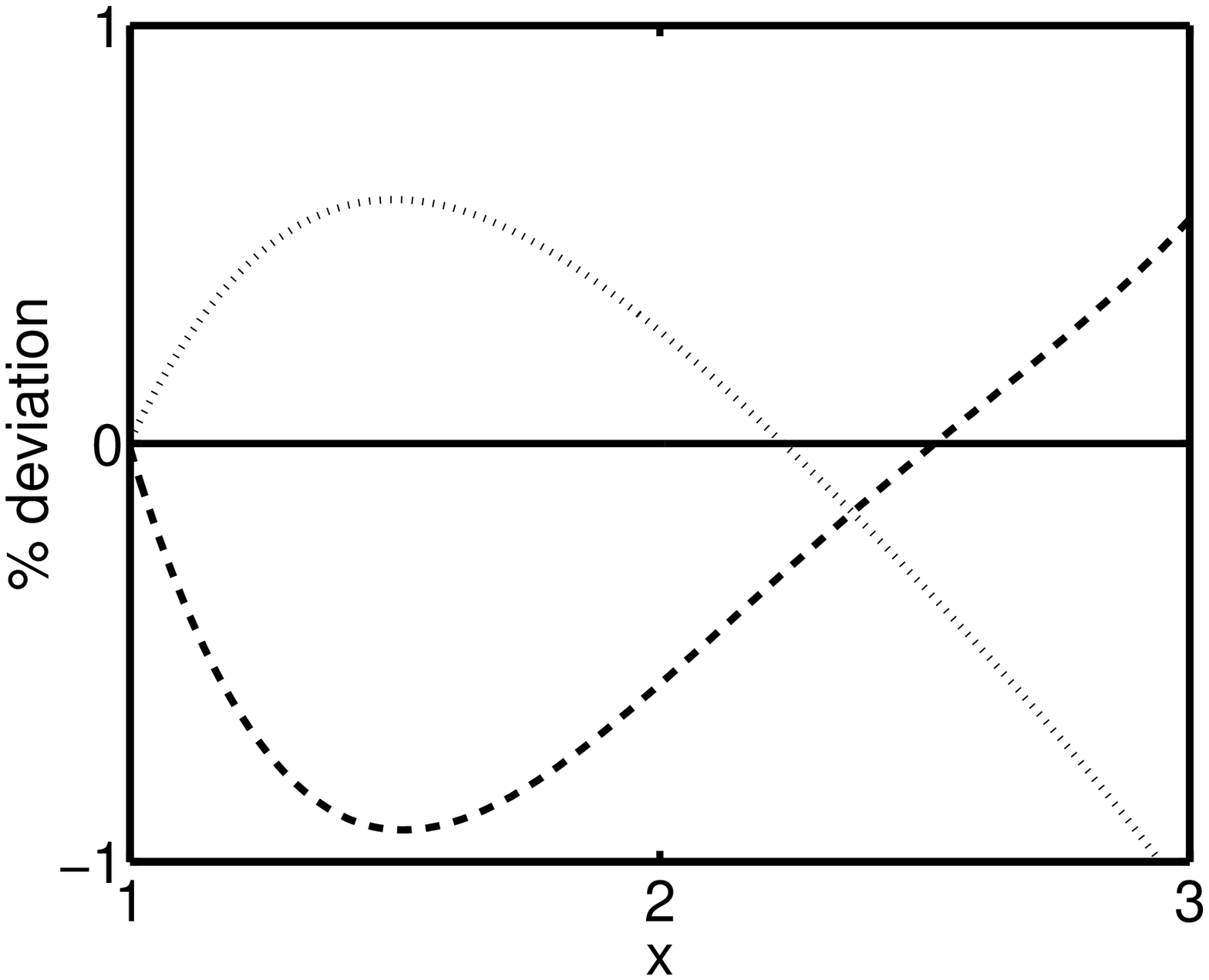,height=35mm}
  \end{center}
\caption{Various exponential potentials (shown on the left), their
           resulting $w_Q$ (middle), and the relative difference in
           $d_L$ (right). The parameters of the
           potentials are listed in table II.}
  \label{exp}
  \end{figure}

\begin{figure}
  \begin{center}
    \epsfig{file=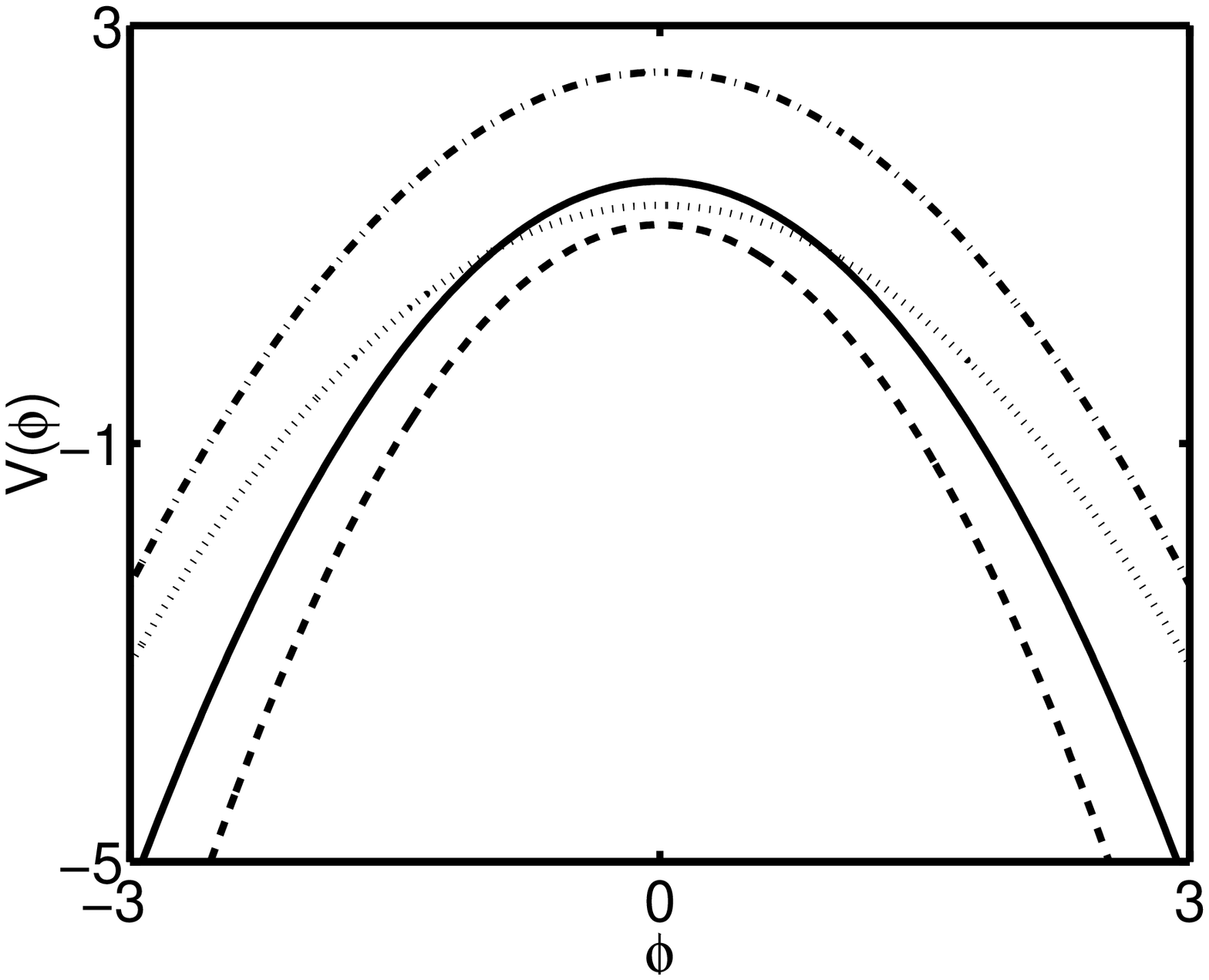,height=35mm}
    \epsfig{file=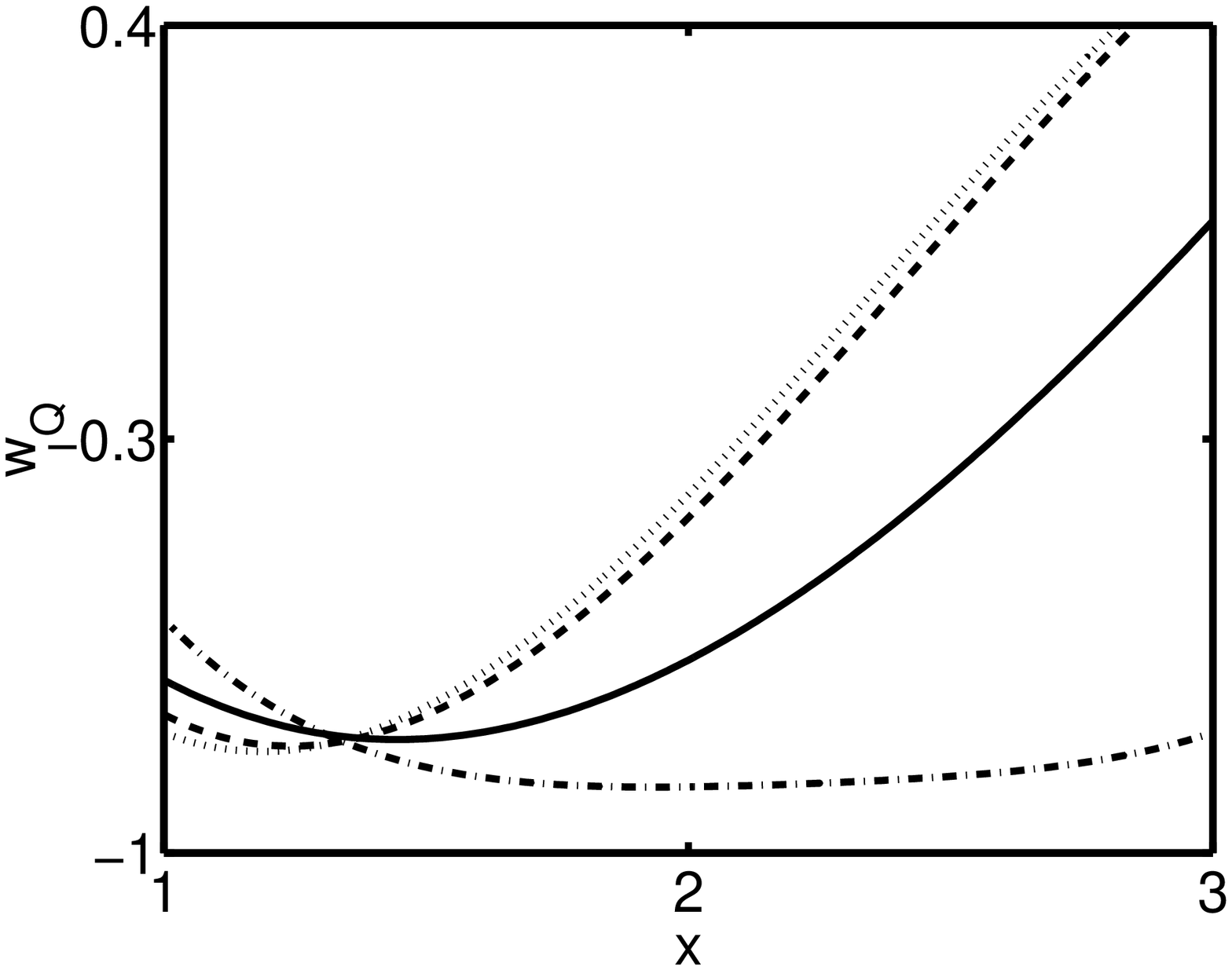,height=35mm}
    \epsfig{file=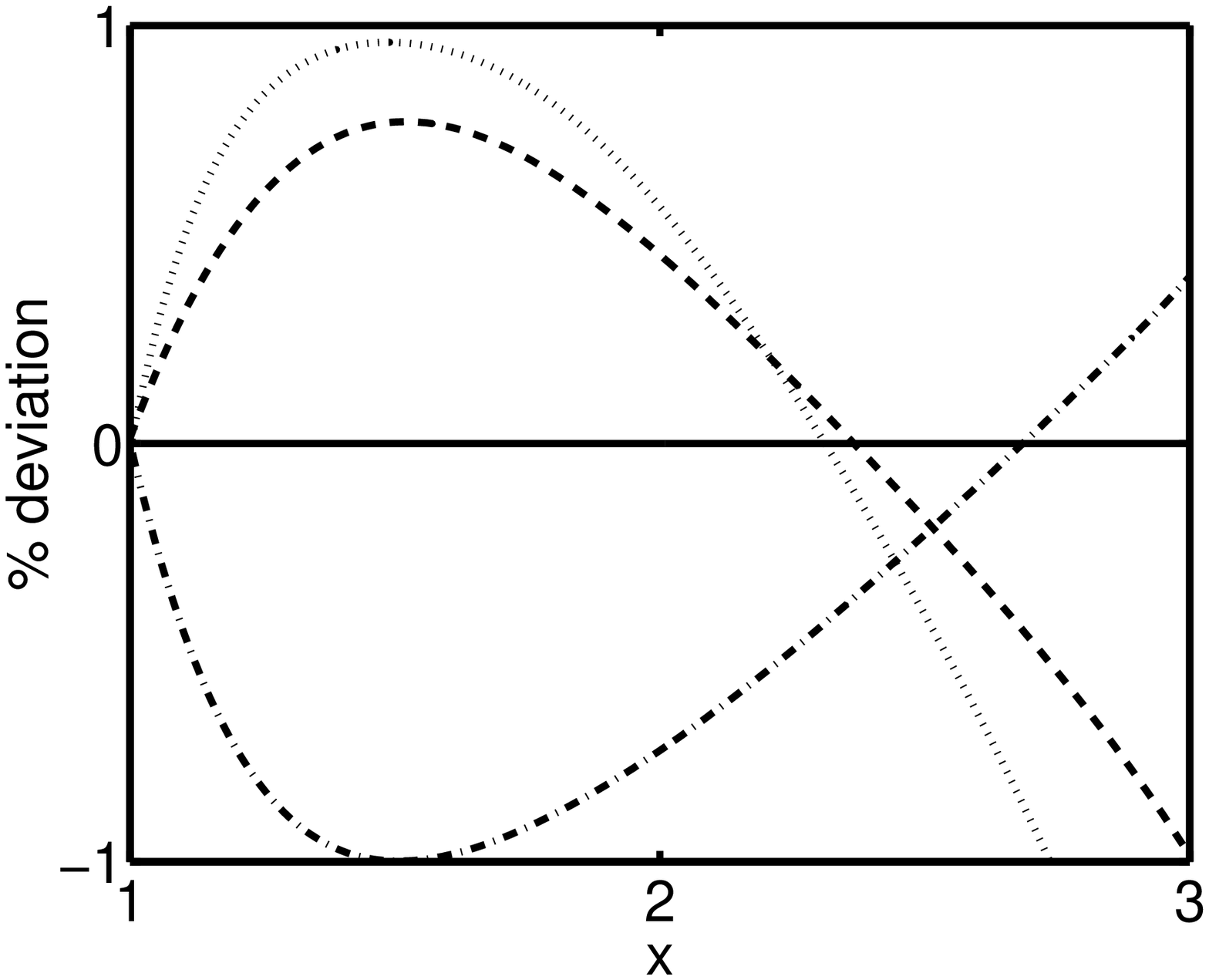,height=35mm}
  \end{center}
  \caption{Various inverted quadratic potentials (shown on the left), their
           resulting $w_Q$ (middle), and the relative difference in
           $d_L$ (right). The parameters of the
           potentials are listed in  table~III.}
  \label{nm}

\end{figure}

Following the same procedure we have analyzed other popular
classes of potentials with two parameters. Figure~\ref{exp} shows
similar results for an exponential potential, $V=Ae^{-B\phi}$. For
this specific potential only some of the solutions of
eqs.~(\ref{algebra1}) and (\ref{algebra2}) can be found
analytically, hence the degeneracy shown is a smaller than the
full degeneracy, as can be seen by the reduced range of allowed
values of $A$ in table II. Figure~\ref{nm} shows the results for
an inverted quadratic potential, $V=-\frac{1}{2}m^2\phi^2+v_0$.

As can be seen from the figures, the deviation in $d_L$ tends to
peak at low redshift, about $z=0.4$. This results from  the
following reason: consider two models whose EOS cross at $z=z^*$.
For redshifts $z<z^*$, the different $w_Q$ of the models imply
different rates of expansion. The model whose $w_Q$ is more
negative  will expand faster, making the $d_L$'s of the two models
diverge. In the range $z>z^*$ the trends reverse, making their
$d_L$'s converge. If this were the only source of degeneracy among
models, it would have been useful to focus measurements in this
range of redshifts, $z\sim 0.4$. Unfortunately a second
degeneracy (due to the uncertainty in $\Omega_m$) degrades the
extra sensitivity in this region of redshifts.

\begin{table}
\begin{center}
\begin{tabular}{|c|c|c|c|c|}
\hline
  & \bf{Solid (fiducial)}& \bf{Dashed} & \bf{Dotted} & \bf{Dash-dotted} \\
\hline
$\bf{(\frac{m}{H_0})^2}$ & 1.51 & 1.12 & 0.98 & 1.99 \\
\hline
$\bf{\frac{v_0}{m_p^2H_0^2}}$ & 0 & -0.86 & 0.20 & 0.47 \\
\hline
$\bf{\Omega_m}$ & 0.30 & 0.31 & 0.29 & 0.29 \\
\hline
\end{tabular}
\end{center}
\caption{Parameters of quadratic potential plotted in
    Figure~\ref{m+}.}
\end{table}

\begin{table}
\begin{center}
\begin{tabular}{|c|c|c|c|}
\hline
  & \bf{Solid (fiducial)}& \bf{Dashed} & \bf{Dotted} \\
\hline
$\bf{\frac{A}{m_p^2H_0^2}}$ & 4.43 & 4.13 & 4.12 \\
\hline
$\bf{m_pB}$ & 1.78 & 2.40 & 1.36 \\
\hline
$\bf{\Omega_m}$ & 0.30 & 0.30 & 0.29 \\
\hline
\end{tabular}
\end{center}
\caption{Parameters of exponential potential plotted in
Figure~\ref{exp}.}
\end{table}

\begin{table}
\begin{center}
\begin{tabular}{|c|c|c|c|c|}
\hline
  & \bf{Solid (fiducial)}& \bf{Dashed} & \bf{Dotted} & \bf{Dash-dotted} \\
\hline
$\bf{(\frac{m}{H_0})^2}$ & 1.51 & 1.88 & 0.98 & 1.08 \\
\hline
$\bf{\frac{v_0}{m_p^2H_0^2}}$ & 1.51 & 1.09 & 1.28 & 2.55 \\
\hline
$\bf{\Omega_m}$ & 0.30 & 0.29 & 0.29 & 0.31 \\
\hline
\end{tabular}
\end{center}
\caption{Parameters of inverted quadratic potentials plotted in
Figure~\ref{nm}.}
\end{table}

So far we have analyzed the uncertainty in determination of
parameters under the assumption that the form of the potential is
known. Without a theoretical prior on the form of the ``true"
potential  it is important to find whether different classes of
potentials can be distinguished by the data alone. We will
present here only a representative example and not attempt a
systematic analysis to expose the full degeneracy.

\begin{figure}
  \begin{center}
    \epsfig{file=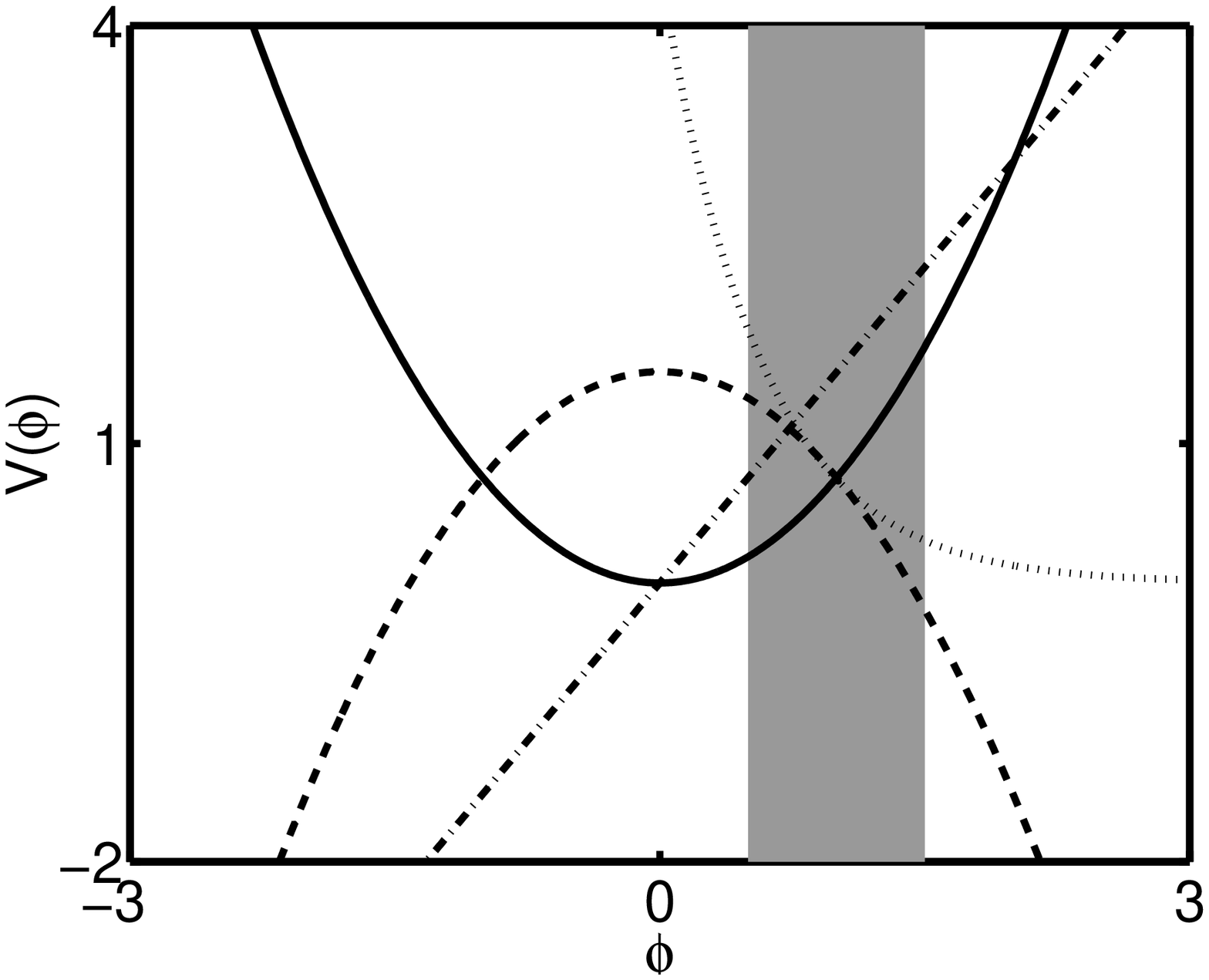,height=45mm}
    \epsfig{file=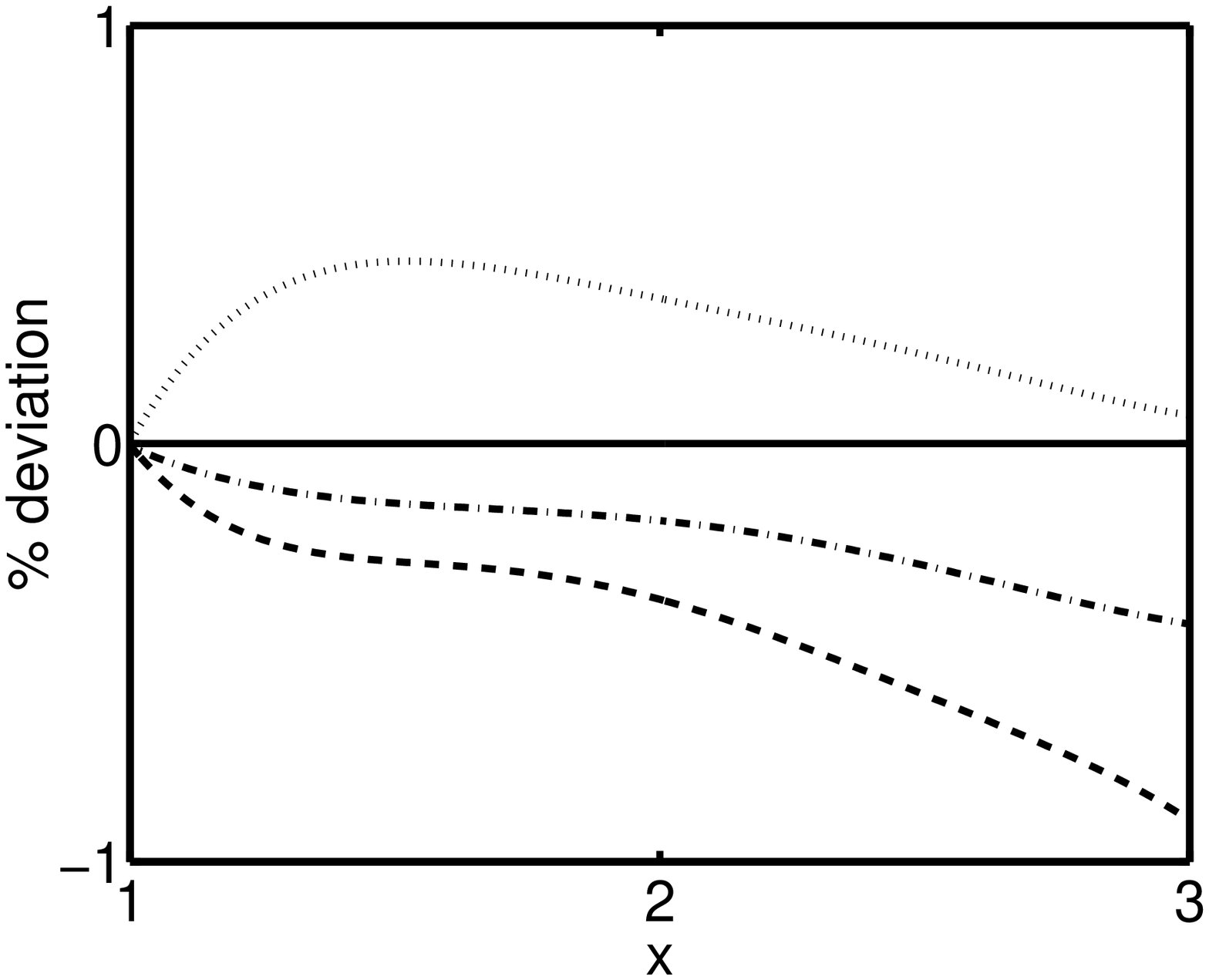,height=45mm}
  \end{center}
  \caption{Various potentials (left panel), and their
           relative $d_L$ differences (right panel).
           The potentials we have used in this example are
           $\frac{1}{2}m^2\phi^2+v_0$ (solid), $Ae^{-B\phi}$ (dotted),
           $-\frac{1}{2}m^2\phi^2+v_0$ (dashed), and $a+b\phi$ (dash-dotted).
           All models have $\Omega_m=0.30$. The shaded area in
           the figure on the left indicates the region in which
           the field moves in the relevant redshift range,
           $0\leq z\leq 2$.}
  \label{mixed}
\end{figure}

In Figure~\ref{mixed} we show an example of four different degenerate
potentials.  All models have $\Omega_m=0.30$. The potentials we have
used in this example are $V=\frac{1}{2}m^2\phi^2+v_0$ (solid),
$V=Ae^{-B\phi}$ (dotted), $V=-\frac{1}{2}m^2\phi^2+v_0$ (dashed), and
$V=a+b\phi$ (dash-dotted). The right panel shows that the relative
differences in their $d_L$'s are less than one percent up to $z=2$.
The left panel shows the potentials.  Although the functional forms
of the potentials are very different and they do not have the same
asymptotic behavior, they are nevertheless indistinguishable.  This
can be understood by observing that the patch of potential that was
probed by the field during the relevant redshift range, (marked in
the figure by the shaded area) is rather small. In this patch,
differences among the various potentials are marginal.

The fact that a small patch of the potential is actually being
probed justifies the use of a simple potential with a small number
of parameters. A fast rolling field would have served as a better
probe of the potential since it would have covered a larger patch
in a given redshift range, but this would have meant a more
positive $w_Q$ which goes against the evidence of accelerated
expansion. Another possible way to enlarge the size of the probed
potential patch  would have been  to measure $d_L$ over a larger
redshift range. The difficulty here is that at deeper redshifts
the contribution of dark energy to the total energy of the
Universe is smaller, and its effects become harder to detect.

\begin{figure}
  \begin{center}
    \epsfig{file=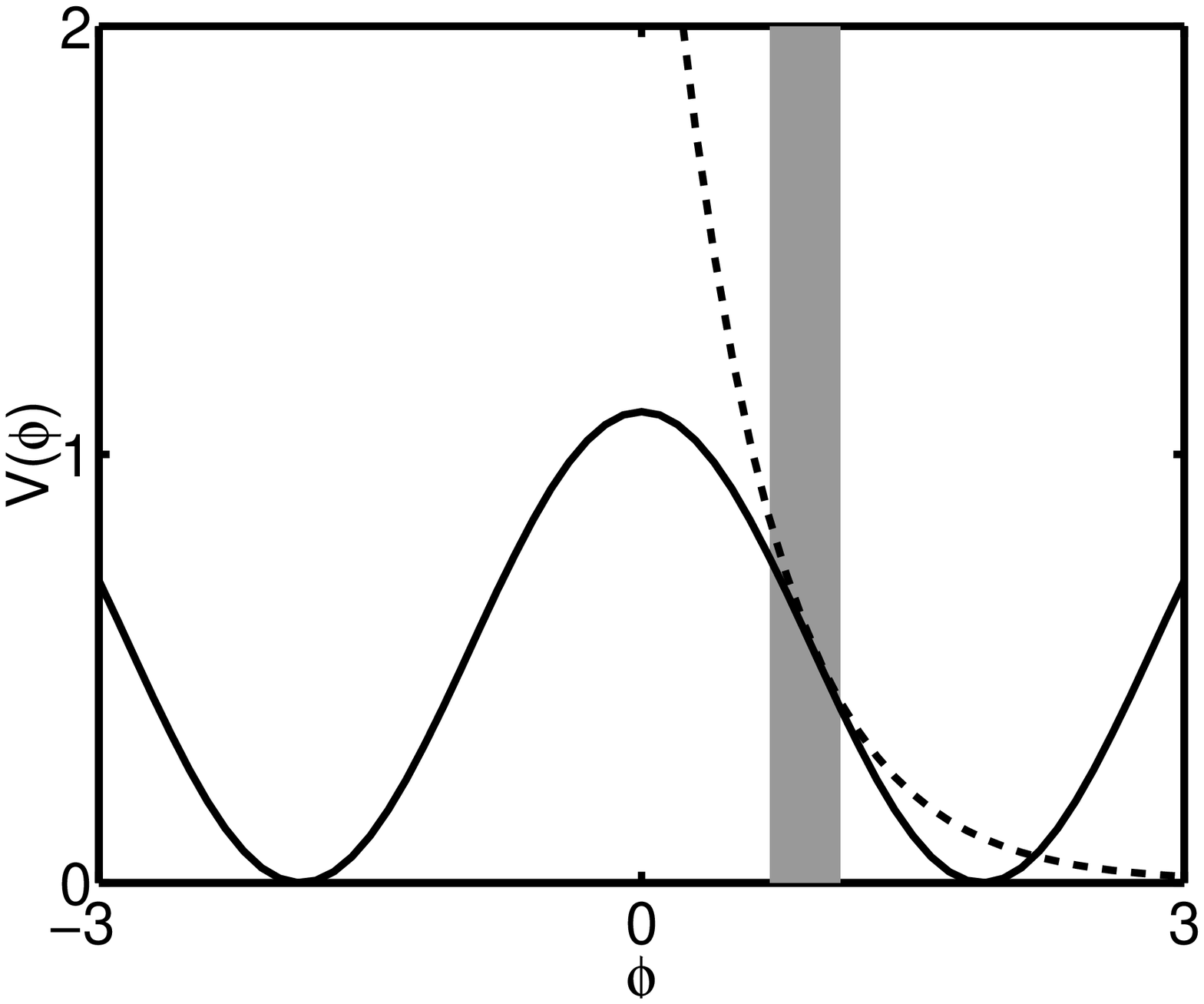,height=45mm}
    \epsfig{file=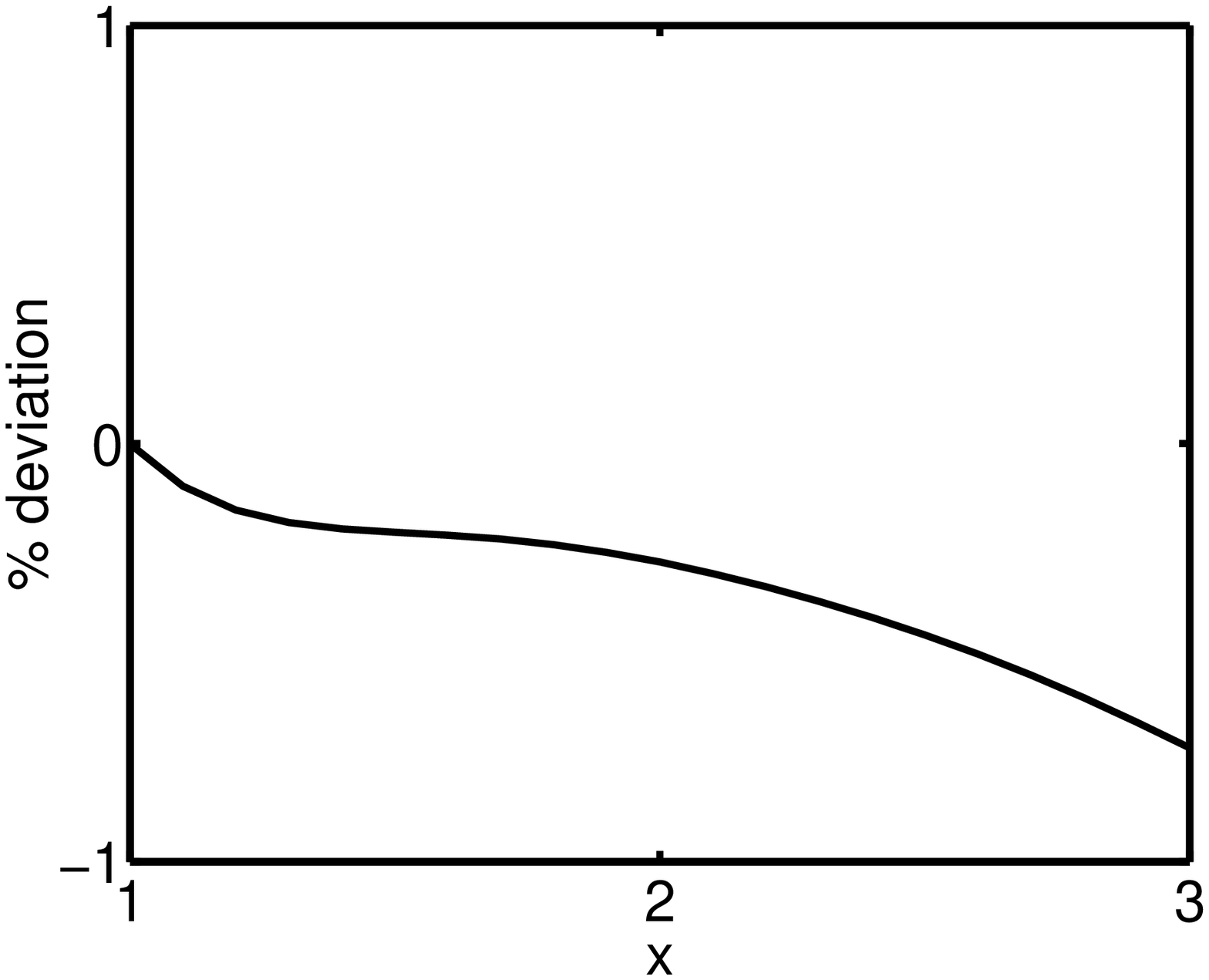,height=45mm}
  \end{center}
  \caption{Two potentials (shown on the left), and their
           relative $d_L$ differences (right).
           The potentials we have used in this example are
           $v=M^4(\cos(\frac{\phi}{f})+1)$ (solid), and
           $Ae^{-B\phi}$ (dashed).
           Both models have $\Omega_m=0.30$. The shaded area in
           the left panel indicates the region in which
           the field moves in the relevant redshift range,
           $0\leq z\leq 2$.}
  \label{weller}
\end{figure}

Our results may seem to disagree with those of \cite{weller},
where it was shown that different forms of potentials could be
distinguished using SNAP-like data alone, but we believe that they
are in fact in full agreement. Our interpretation of the results
of \cite{weller} is that they depend on the choice of specific
parameters and initial conditions for each of the potentials that
makes them distinguishable. This does not mean that for a given
set of data, it will be possible to determine that a specific
class of potentials is preferred over another. It is quite easy to
construct counter examples by choosing the potentials to look
similar in the relevant redshift patch. A representative example
is shown in Fig.~\ref{weller}. The potentials that are used here
are the pseudo-Nambu-Goldstone boson potential,
$V=M^4(\cos(\phi/f)+1)$ (with $M^4/(m_p^2H_0^2)=0.82$ and
$f/m_p=0.6$), and the pure exponential, $V=Ae^{-B\phi}$ (with
$A/(m_p^2H_0^2)=4.43$ and $m_pB=1.78$).

\section{Considering CMB}

An additional measurement that can possibly probe the dark energy
EOS and the scalar field potential is that of the  cosmic
microwave background (CMB) anisotropy. This measurement is
expected to be improved with the upcoming MAP and
 PLANCK missions.

Using the CMB measurements to probe scalar field models requires
additional theoretical assumptions about the evolution in the
range from $z\sim 2$ to $z\sim 1000$. For example, that the
potentials are ``tracker" potentials \cite{Zlatev}, or that the
scalar field has negligible influence above some redshift (say,
$z=2$).

It is well-known that the EOS cannot be determined using CMB
anisotropy data alone due to degeneracies. A thorough analysis was
done in \cite{huey}, who concluded that this degeneracy persists
after combining CMB and SN measurements. In \cite{m2} it was
shown that the inclusion of CMB constraints  in the analysis of
the dark energy EOS does improve the resolution somewhat, but not
significantly.

In a simple minded estimate here, we would like to show that under
reasonable assumptions, CMB measurements will not help to
significantly constrain the potential either. Rather than
performing an extensive numerical search, as in
\cite{m2,copeland,fhlt}, we use a different strategy: we
treat CMB measurements as effectively providing one additional
point on the Hubble diagram at the last scattering redshift
$z_{ls}\sim 1000$. The method
cannot provide numerically accurate results, but it does highlight
the theoretical degeneracy that limits the
ability of CMB measurements to constrain the parameters and functional
form of the potential.

Treating the CMB measurements as effectively  one
additional point on the Hubble diagram at redshift $z\sim 1000$
can be implemented in a simple way because the angular distance
and the luminosity distance are related to each other
 $d_L=(1+z)^2 d_A.$ The observed angular
size of any feature in the CMB, $\theta$, is related to its
physical size $d$  by the angular distance $d_A$ to last
scattering surface, $\theta d_A=d$. The CMB spectrum yields a
series of acoustic peaks, located at $l_n=\frac{n\pi}{S}d_A$, $n$
being an integer, and $S$ is the sound horizon at last scattering
surface. We are going to ignore uncertainties in the sound
horizon $S$, which depends on the composition of the universe at
last scattering surface.

To estimate the measurement accuracy of the CMB point on the
Hubble diagram we may treat the position of the first peak $l_1$,
as a direct measurement of the angular distance, therefore the
relative errors in $l_1$ and in the angular distance are equal
$\frac{\Delta l_1}{l_1}=\frac{\Delta d_A}{d_A}$. The position of
the first peak is already measured rather well, to about $3\%$ at
the $1\sigma$ level (see, for example, \cite{hu}). This is
expected to be somewhat improved by MAP and PLANCK, so we may
expect eventually a $3\sigma$ error in the few percent range.

Another way of estimating the accuracy of the CMB point is to use
the results of \cite{m2}, where it was shown that for models
which have a linear $w_Q$ for $z<2$ and $w_Q=const.$ from $z=2$
to last scattering surface, a derivative as large as
$w_1=\pm\frac{1}{6}$ could not be distinguished at the $3\sigma$
level, assuming full sky coverage and cosmic variance limited
measurement. This means that the most accurate CMB measurements
do not distinguish models which lead to a difference of (+1.6\%,
-3.2\%) in their $d_L$'s. Thus a $3\sigma$ measurement error
estimate of a few percent in $d_L$ seems reasonable.

According to the preceding discussion we may define models to be
indistinguishable by the CMB if their $d_L$ difference is less
than a few percent at $z_{ls}$. We may check now whether this
imposes additional constraints on our models. We find that it
does, but not to a significant degree.

Since our toy models are not strictly chosen as tracker models,
evolving the equations of motion backwards from $z\sim 0.3$ up to
$z\sim 1000$ gives nonsensical results, the dark energy
typically becomes dominant, with $w_Q$ approaching 1. In order to
to avoid such undesirable and observationally excluded behavior,
the potential needs to be changed in the regions that are probed
by the field in the redshift range $z\gg 1$ such that the kinetic
energy remains acceptably small, so that $w_Q$ remains negative
and $\Omega_Q$ becomes sub-dominant.

Instead of building the potential piece-wise, we have chosen
another approach in the spirit of \cite{m2}: we let the equations
of motion evolve until $z=2$ (the redshift range relevant to SN
measurement), and for $z>2$ we put by hand $w_Q=0$ for all models.
We then calculate numerically the relative differences in $d_L$ at
$z=1000$. Our cutoff procedure changes the $z$-dependence of $w_Q$
at a value of $z$ where the dark energy density is negligible,
which makes the details of the $z$-dependent cutoff unimportant.
We expect other cutoff procedures to give similar results.

We find that the relative differences in $d_L$ for our models are
within a few percent, in agreement with the semi-analytic
argument that we present shortly. For the potentials shown in
Fig.~\ref{mixed}, we find $\frac{\Delta d_L}{d_L}=$ $-0.9$\% for
the exponential potential, $-3.4$\% for the inverted quadratic
potential, and $-1.6$\% for the linear potential. As explained
previously this is just at the limit of CMB resolution. For the
quadratic potential examples in Fig.~\ref{m+} we find the
following $d_L$ differences from the fiducial, dashed: $2.5$\%,
dot-dashed: $-6.2$\%, dotted: $8.0$\%, for the exponential
potentials of Fig.~\ref{exp} we find: dashed: $4.5$\%, dotted:
$6.8$\%, and for the inverted quadratic potential of
Fig.~\ref{nm} we find dashed: $-7.8$\%, dotted: $-8.1$\%, and
dot-dashed: $5.1$\%.

For most of the models we see that CMB is expected to further
constrain the parameters if the form of the potential is known.
The larger $d_L$ differences are for models whose differences at
$z\sim 2$ are larger, in agreement with the following
semi-analytic argument. Given the approximate nature of our
measurement accuracy estimate, the large number of additional
sources of degeneracy that we have neglected, and the additional
theoretical assumptions that go into the analysis with the CMB
point added it is not possible, and we believe that it is  not
necessary to determine in a more quantitative way the amount by
which the degeneracy is reduced.

We may estimate  in a rough semi-analytic way the relative errors
in $d_L$ at $z\sim 1000$ as follows. Luminosity distance at
$1+z_{ls}=x_{ls}\approx 1000$ for models with $w_Q(x>3)=0$ is
given by \cite{m2},
 \bea
 \label{dl1}
    d_L(x_{ls})&=&x_{ls}\sqrt{1+g}\int_1^{x_{ls}}\frac{dx}{x^{3/2}}
        \left( g+exp(3\int_1^x w_Q\frac{dy}{y}) \right)^{-1/2}
    \nonumber \\
    &=&x_{ls}\sqrt{1+g}\int_1^{3}\frac{dx}{x^{3/2}}
        \left( g+exp(3\int_1^x w_Q\frac{dy}{y}) \right)^{-1/2}  \nonumber \\
      &+&x_{ls}\sqrt{1+g}\int_3^{x_{ls}}\frac{dx}{x^{3/2}}
        \left( g+exp(3\int_1^x w_Q\frac{dy}{y}) \right)^{-1/2} \nonumber \\
    &=&\frac{x_{ls}}{3} d_L(3)+
        x_{ls}\sqrt{1+g}\int_3^{x_{ls}}\frac{dx}{x^{3/2}}
        \left( g+exp(3\int_1^3 w_Q\frac{dy}{y}) \right)^{-1/2} \nonumber \\
    &=& \frac{x_{ls}}{3} d_L(3)+
        2x_{ls}\left(\frac{1}{\sqrt{3}}-\frac{1}{\sqrt{x_{ls}}} \right)
        \left( \frac{1+g}{g+exp(3\int_1^3 w_Q\frac{dy}{y})}
        \right)^{1/2},
 \eea
where $g$ is the current ratio of matter to dark energy densities,
and we have set $H_0$ to unity. The second term can be expressed in
terms of the luminosity distance at $x=3$ and its derivative: 
  \begin{equation}
     \left( \frac{1+g}{g+exp(3\int_1^3 w_Q\frac{dy}{y})} \right)^{1/2}=
        3^{3/2}\frac{d}{dx}\left(\frac{d_L}{x} \right)_{x=3},
 \end{equation}
so,
\begin{equation}
 d_L(x_{ls})
     \simeq -\frac{x_{ls}}{3} d_L(3)+
        2x_{ls} d_L'(3).
\end{equation}

Now we would like to examine the difference in $d_L(x_{ls})$ for
two models. For a simple and rough error estimate, we may use
$d_L'(3)= c_1 \frac{d_L(3)}{3}$ and $\Delta d_L'(3)=c_2
\frac{\Delta d_L(3)}{3}$ with $c_1$, $c_2$  numerical
coefficients of order unity. So finally we obtain
 \begin{equation}
 \left(\frac{\Delta d_L}{d_L}\right)_{x_{ls}}\simeq \frac{2c_2-1}{2c_1-1}
 \left(\frac{\Delta d_L}{d_L}\right)_{x=3}.
 \end{equation}
Note that the relative $d_L$ error at $x_{ls}$ is independent
of $x_{ls}$. Since we consider models whose maximal
relative error at $z=2$ is one percent,
they will have a relative error in $d_L$ of about a
few percent at $z_{ls}$, depending on the values of $c_1$ and $c_2$.
This estimate agrees with the numerical examples that we have listed above.

We conclude that once physically reasonable constraints are
imposed on the potential, CMB measurements do not significantly
constrain the parameter space of the potentials, although they
are expected to reduce it somewhat. Our simple minded analysis
strengthens the case first made in \cite{huey} where it was shown
that only some average EOS can be measured, and agrees with
\cite{linder,fhlt}.

\section{Conclusions}

We have found that it is not possible to obtain  precise
quantitative estimates  for parameters of scalar field models of
dark energy from data alone beyond the obvious order of magnitude
estimates. This is  due to theoretical degeneracies, which would
persist even with expected future data from the most accurate SN
and CMB measurements.

Theoretical prior knowledge or assumptions on the form of the
potential and the field's initial conditions (preferably leaving
a total of just two free parameters) may allow a more quantitative
determination. For example, assuming that the field is at rest at
the bottom of the potential is equivalent to having a pure
cosmological constant. In this case the magnitude of the
cosmological constant can be determined with accurate data to
within a few percent.

\section{Acknowledgements}

I.M. gratefully acknowledges support from the Clore foundation.


\end{document}